\documentclass[
]{ceurart}

\usepackage{listings}
\begin{document}

\copyrightyear{2026}
\copyrightclause{Copyright for this paper by its authors.
  Use permitted under Creative Commons License Attribution 4.0
  International (CC BY 4.0).}

\conference{Bridge Over Troubled Water: Aligning Commercial Incentives With Ethical Design Practice To Combat Deceptive Patterns. Workshop at the 2026 CHI Conference on Human Factors in Computing Systems (CHI EA '26), April 13--17, 2026, Barcelona, Spain.}

\title{Dark Patterns in Indian Quick Commerce Apps: A Student Perspective}

\author[1]{Tanish Taneja}

\author[1]{Arihant Tripathy}

\author[1]{Nimmi Rangaswamy}

\address[1]{International Institute of Information Technology Hyderabad,
Gachibowli, Hyderabad, Telangana 500032, India}
\begin{abstract}
  As quick commerce (Q-Commerce) platforms in India redefine urban consumption, the use of deceptive design dark patterns to inflate order values has become a systemic concern. This paper investigates the `Awareness-Action Gap' among Indian university students, a demographic characterized by high digital fluency yet significant financial constraints. Using a qualitative approach with 16 participants, we explore how temporal pressures and convenience-driven architectures override price sensitivity. Our findings reveal that while students recognize manipulative UI tactics, they frequently succumb to them due to induced cognitive load and the normalization of deceptive marketing as a `price of capitalism'. We conclude by suggesting value-sensitive design alternatives to align commercial incentives with user autonomy in the Global South.
\end{abstract}


\maketitle

\section{Introduction}
The Indian digital retail landscape has recently witnessed a seismic shift from traditional e-commerce to `Quick Commerce' (Q-Commerce). The rise of relatively new platforms such as Zepto, Blinkit (Zomato) and Instamart (Swiggy) \cite{qcom-gupta}  have redefined urban consumption by promising delivery timelines as quick as 10 minutes. This hyper-local model leverages `Dark Stores' \cite{qcom-goyal} and advanced logistics to cater to immediate needs. However, this convenience comes at a psychological cost. To sustain the high operational expenses of ultra-fast delivery, these platforms are forced to rely heavily on maximizing `Average Order Value' (AOV) and order frequency.

Existing HCI literature defines `Dark Patterns' as interface designs that trick users into actions they would not otherwise take \cite{gray2018}. While traditional e-commerce allows users time to compare prices and deliberate, Q-Commerce is structurally designed to eliminate deliberation. The promised instant gratification by these platforms naturally creates a high-pressure environment where speed comes before scrutiny.

For millions of urban Indians, these Quick Commerce apps have transitioned from luxury services to essential infrastructure for `just-in-time' living. Despite the government's recent Guidelines for Prevention and Regulation of Dark Patterns \cite{ccpa2023}, users are frequently unable to identify and ignore manipulative tactics. The promise of 10-minute delivery offers such immense utility that users often choose to succumb to the pattern rather than face the friction of traditional alternatives. This suggests that the rapid growth of Quick Commerce has made convenience a more valuable currency than transparency.

This paper investigates a specific demographic of Indian University Students. Students represent a unique intersection of high digital literacy but also high financial vulnerability. While they might be more familiar with the latest technology to be able to recognize manipulative UI tactics, yet they are also resource-constrained consumers who should theoretically be highly price-sensitive.

Our central hypothesis is that \textbf{participants can identify obvious dark patterns (awareness) but still succumb to them (action) due to the induced cognitive load of `Time Scarcity' and `Convenience'.} Through structured user interviews, we try to understand how these various elements in the environment of quick commerce interact with the users to affect their financial choices and nudge them towards prioritizing immediate relief over long-term savings.

By focusing on the `Awareness-Action Gap', we aim to move beyond listing dark patterns to understanding why they work on a population that can, in theory, see through them. We ask three concrete questions. First, do participants actually recognise manipulation as it happens, or only in hindsight and does that timing matter? Second, when they do recognise it, what rationalisations make the added cost feel acceptable to people who are otherwise budget-conscious? Third, and most critically: does awareness of a deceptive flow; an obscured cancellation path, an artificial countdown, a fee buried at checkout realistically translate into any real resistance, or does the structure of Q-Commerce make opting out feel more costly than giving in?

\section{Related Work}
\subsection{Dark Patterns and their Taxonomy}
The concept of Dark Patterns was first introduced by practitioner Harry Brignull in 2010 to describe user interface (UI) design choices that benefit an online service by coercing, steering, or deceiving users into making unintended decisions \cite{brignull-website}. Initially, it started as a practitioner-led concept but since then, it has been rigorously defined in Human-Computer Interaction (HCI) literature. Unlike the pattern of `Nudging' \cite{thaler2008nudge} which aims to guide users toward beneficial choices without restricting options, other dark patterns are distinct in their reliance on deception and information asymmetry to subvert the user autonomy.

This definition has been further formalized in the context of e-commerce and used to describe dark patterns as design choices that modify the architecture to manipulate user behavior against their best interests \cite{dark-patterns-at-scale}. More recently, definitions have expanded to also include the legal dimension and argue that these patterns operate by exploiting cognitive biases to such a degree that they effectively deprive users of informed consent, rendering the resulting interactions legally dubious \cite{luguri2021shining}.

To fully understand the prevalence of deceptive design in modern interfaces, it is necessary to trace their origins beyond the digital age and look at them because of three distinct historical trends converging over the last several decades \cite{Narayanan2020DarkPP}. 1. \textbf{Deceptive Retail Practices}: Traditional retail manipulation tactics such as psychological pricing (\$9.99 vs \$10.00), bait-and-switch advertising and false store closing had already demonstrated that subtle changes in presentation could profitably alter consumer behavior. 2. \textbf{The Nudge Revolution}: The second force was the rise of behavioral economics in the 1970s as scholars introduced choice architecture to help people make better decisions through gentle nudges \cite{thaler2008nudge}. 3. \textbf{Growth Hacking Culture}: The growth hacking mindset of the 1990s and 2000s emerged from tactics like Hotmail's promotional signatures and created a culture which prioritized rapid adoption above all else. Aggressive A/B testing and metric-driven optimization turned user manipulation into a standard engineering practice \cite{juneja2025}.

Today, these forces have taken the form of a sophisticated ecosystem where companies can integrate manipulative tactics directly into code, targeting financial resources, data, and attention with high precision.

While these are initial definitions, there are significant works in establishing the foundational taxonomy widely used in HCI \cite{gray2024, gray2018, dark-patterns-at-scale}. As a result, the classification of dark patterns has evolved significantly since their initial identification. To encompass these wide range of dark pattern categories and subtypes, we present a summary in Table \ref{tab:taxonomy}.

While foundational ontologies \cite{gray2018, gray2024} provide a robust theoretical lens for categorizing deceptive design, Table \ref{tab:taxonomy} employs a synthesized taxonomy that aggregates these patterns into clusters aligned with the 2023 CCPA Guidelines \cite{ccpa2023}. This reconfiguration also maps the academic definitions (e.g., `forced enrollment' or `hidden costs') onto the specific prohibited practices recognized by Indian regulators and we try provide a more direct bridge between HCI research and current policy enforcement in the Indian ecosystem.

\begin{table}[t!]
  \centering
  \resizebox{\textwidth}{!}{%
    \begin{tabular}{|l|l|l|}
      \hline
      \textbf{Category}      & \textbf{Sub-categories}                & \textbf{Examples (NOT exhaustive)}                             \\ \hline
      Nagging                & Persistent, Intermittent               & Recurring membership prompts such as seasonal reminders        \\ \hline
      Obstruction            & Roach Motel, Hard to Cancel            & Complex cancellation flow, Misleading fees                     \\ \hline
      Sneaking               & Bait and Switch, Hidden Costs          & Weird default settings, Last minute charges                    \\ \hline
      Forced Action          & Mandatory Sharing, Forced Registration & Intrusive App Permissions, Lack of features without an account \\ \hline
      Social Proof           & Fake Activity, Manufactured Scarcity   & False item counts and buying trends                            \\ \hline
      Interface Interference & Hidden Information, Preselection       & Pre-selected Checkboxes, Buried Terms and Conditions           \\ \hline
      Urgency                & Limited Time Offers, Countdown Timers  & Artificial Deadlines, Misleading Claims                        \\ \hline
      Misdirection           & Trick Questions, Visual Interference   & Double Negatives, Targeted Color Schemes                       \\ \hline
      Privacy Zuckering      & Friend Spam, Data Exploitation         & Ambiguous Data Collection and Sharing, Intrusive Permissions   \\ \hline
    \end{tabular}%
  }
  \caption{Summarized Taxonomy for Common Dark Patterns}
  \label{tab:taxonomy}
\end{table}

\subsection{Dark Patterns in Online Shopping Apps}
The commercial incentive to maximize conversion rates has made e-commerce the ideal environment for deceptive design. The largest empirical study in this domain \cite{dark-patterns-at-scale} analyzes over 50,000 product pages across 11,000 shopping websites. As a result, it uncovers that dark patterns are not only edge cases but systematic features of the online retail ecosystem, identifying 1,818 distinct instances across 15 types. The key findings from this research established that shopping platforms primarily utilize Information Asymmetry to manipulate purchase decisions. The most dominant categories of dark patterns that were identified are: \textbf{False Urgency} - Countdown timers that reset on page reload, creating artificial time pressure; \textbf{Activity Messages} - Notifications (e.g., `Jane from Ohio just bought this') used as social proof to trigger the bandwagon effect; \textbf{Sneaking} - The non-consensual addition of items or fees into the shopping cart, a practice found even in established retailers, not just obscure sites \cite{dark-patterns-at-scale}.

As e-commerce also shifts to mobile applications, the nature of user vulnerability changes yet again. In fact, it is argued that mobile apps present a more dangerous context for dark patterns due to screen constraints and interaction modalities \cite{chen2023unveiling}. While web browsers allow users to inspect elements and compare across tabs, mobile apps operate with limited screen real estate and make Interface Interference (hiding information) significantly easier to incorporate while also harder to detect \cite{chen-2024, chen2023unveiling}.

Additionally, mobile dark patterns often rely on visual misdirection (such as making the Cancel button distinctively smaller or lower-contrast than the Accept button) to exploit errors and the user's tendency to skim through rather than read in detail on mobile devices \cite{chen-2024}. These mobile patterns are also dynamic and appear only after specific interaction sequences (e.g., scrolling to the bottom of a cart), meaning that they go undetected in static analyses.

Recent literature further expands the understanding of dark patterns beyond static UI elements to include temporal dark patterns that weaponize time and flow. A famous example, called the Iliad Flow, emerged while analyzing the sequence of Amazon Prime cancellation process \cite{gray2025iliad}. It argues that obstruction is not just a barrier but also a temporal experience. Platforms choose to intentionally lengthen the time for detrimental actions (cancellation) while shortening it for beneficial actions (one-click purchase) and introduce temporal friction to exhaust the user's cognitive resources and try to get them to resign.

Even in live-stream commerce on platforms like TikTok and Taobao, malicious selling strategies (such as real-time countdowns and verbal pressure from streamers) have been identified to create a state of panic buying. Such planned contexts are done to inhibit rational decision making and force users to ignore price details and focus solely on securing the item before time runs out \cite{wu2022malicious}.

While much of the early dark pattern literature focuses on web-based e-commerce, recent works also emphasize the unique risks posed by mobile applications. Mobile apps are shown to leverage `interactional' dark patterns that exploit the user's need for speed and limited screen real estate \cite{geronimo-etal}. Furthermore, it has been highlighted that the shift from desktop to mobile introduces modality-specific deceptions, where haptic feedback and swipe-based architectures can bypass a user's critical reflection \cite{gunawan-etal}. Finally, large-scale analysis has further confirmed that these patterns are pervasive in top-tier mobile applications \cite{chen-etal}, suggesting that the quick-commerce ecosystem in India is part of a broader, mobile-first trend of algorithmic manipulation.

This context is particularly relevant when we start to discuss Quick Commerce later since the mobile-first and time-critical nature of interactions only amplifies the efficacy of these patterns.

\subsection{Dark Patterns in the Indian Context}
The design of dark patterns is also deeply influenced by cultural contexts, something that can often be overlooked while making universal assumptions in Human-Computer Interaction (HCI) research. As noted in \cite{juneja2025}, the majority of existing literature has centered on Western contexts, which prioritize individualism and exhibit higher skepticism toward manipulative tactics \cite{gray2018}. This creates a critical gap, as cultural values significantly shape how users interpret design cues.

In the Indian context, specific cultural dimensions identified by Hofstede, namely High Power Distance and Collectivism, amplify the effectiveness of certain dark patterns \cite{juneja2025}. It is further suggested that in high power-distance societies, users are more likely to trust authority cues, making `Authority Endorsements' (e.g., `Recommended by Experts') disproportionately effective. The effectiveness of social proof in the Indian context is also further amplified by cultural dimensions such as collectivism and high power distance, as suggested by prior HCI research in the Global South \cite{juneja2025}. In such contexts, community validation and authority cues have been seen to become primary drivers in digital decision-making.

India also presents another challenge due to its rapid digitization relative to digital literacy. With over 700 million internet users, a significant portion is made up of `first-generation' users accessing the web exclusively via mobile apps \cite{juneja2025}. In this study, we consider digital literacy to be a high degree of operational fluency and the ability to navigate complex, multi-step mobile interfaces. However, this does not directly guarantee critical design literacy and the ability to understand the underlying choice architectures.

The existing `Awareness-Action Gap' is also further widened by contextual factors such as convenience and temporal urgency inherent to urban living conditions \cite{chugh2024unpacking}. Also, we note that this observed phenomenon of the `Awareness-Action Gap' aligns with prior work as well \cite{bongard2021}, where it was observed that users often do feel `definitely manipulated' even when they are fully aware of the patterns, describing the state as `ridiculous' yet inescapable.

In the analysis of the Indian digital economy, it is also important to emphasize that dark patterns in India do not merely cause annoyance but also erode autonomy, even in critical sectors like finance and healthcare \cite{raj2025safeguarding}. For instance, patterns of the form of `Forced Action' and `Nagging' are frequently deployed in sensitive applications to steer users into financial products they do not need, exploiting the information asymmetry inherent in the Indian market \cite{chugh2024unpacking}.

Unlike the established GDPR framework in Europe, India's regulatory response is still in a nascent stage but also rapidly evolving. We look at this in more detail in Section \ref{relatedwork-legal}.

\subsection{Quick Commerce Landscape in India}
Since the last 4--5 years, the retail landscape in India has undergone a massive shift with the emergence of `Quick Commerce' (Q-commerce). Unlike traditional e-commerce, which focuses on wide selection and standard delivery timelines (1--3 days), Q-commerce is defined by its promise of hyper-local delivery within minutes, typically ranging from 10 to 30 minutes \cite{qcom-goyal}. As noted in \cite{qcom-gupta}, this model has evolved as a `disruptive force' that fundamentally redefined consumer expectations regarding speed and convenience.

The rapid adoption of Q-commerce in India is credited to certain socio-economic factors. \cite{qcom-singh} identifies that the surge is driven by an overall increased smartphone penetration and a behavioral shift towards instant gratification, where the reduced time-to-deliver is valued over price savings. This is particularly prevalent in dense urban centers. \cite{qcom-goyal} argues that evolving urban lifestyles, characterized by `time scarcity', have transformed convenience from a luxury into a necessity. As a result, platforms like Blinkit, Zepto, and Swiggy Instamart have successfully marketed themselves to create a behavioral loop where patience is eroded by the availability of instant solutions \cite{qcom-singh}.

Even though consumer adoption has been robust, the economic viability of Q-commerce remains a critical challenge as the initial phase of the industry was driven up through heavy venture capital infusion and deep discounting to acquire users \cite{qcom-ganpathy}. However, the sector is now facing a situation where the focus must shift from valuation to profitability. It can be said that generating revenue solely from delivery charges is insufficient to cover the high operational costs of dark stores. This creates an intense pressure on platforms to increase the Average Order Value (AOV) and minimize customer churn \cite{qcom-ganpathy}.

This transition which sees the sector go from exponential growth at any cost to survival via profitability provides the crucial context for this study. The aggressive `dark pattern' design tactics observed in these apps may not be accidental, but rather calculated attempts to solve the unit economics problem by artificially inflating order values and purchase frequency.

\subsection{Legal Guidelines against Dark Patterns}
\label{relatedwork-legal}
The regulation of deceptive design has historically been led by the European Union (EU) and the United States (USA). In the EU, dark patterns are primarily tackled through a mix of data protection and consumer laws. The \textbf{General Data Protection Regulation} (GDPR) and the \textbf{Unfair Commercial Practices Directive} (UCPD) collectively target design choices that impair user autonomy or cause `material and non-material harm' \cite{santos-legal}. Specifically, the EU focuses on patterns that manipulate consent or distort economic behavior. Similarly, in the United States, the \textbf{Federal Trade Commission} (FTC) relies on Section 5 of the FTC Act to prosecute unfair or deceptive acts \cite{mamidwar-legal}.

More recently, India has emerged as a proactive regulator in the Global South. The pivotal introduction of the `Guidelines for Prevention and Regulation of Dark Patterns, 2023' by the \textbf{Central Consumer Protection Authority} (CCPA) \cite{ccpa2023} has been a big step towards regulation \cite{awasthi2025dark}. These guidelines explicitly classify 13 specific practices (most of those similar to the taxonomy defined earlier) as unfair trade practices.

Despite these robust definitions, a significant gap remains between policy and practice. It has been observed that while the laws exist, there is little to no enforcement effort to effectively curb these practices in real-time, particularly in the fast-moving e-commerce ecosystem \cite{mamidwar-legal}. A massive gap still remains between the laws and the apps as platforms continue to exploit loopholes in the guidelines through `interface interference' that is technically legal but functionally deceptive \cite{kumar2025dark}.

This regulatory lag provides more importance for this research as India's digital market keeps expanding significantly faster than the regulatory frameworks can be enforced \cite{juneja2025}. Moreover, investigating the interaction of dark patterns with the resource-constrained student population offers practical value by comparing the guidelines with the realities present in these apps. Theoretically, it moves UX research beyond the frameworks established by the GDPR (in the EU) and FTC (in the US) and acknowledges that legal compliance in India must account for distinct cultural and economic vulnerabilities.
\section{Methodology}

To test our hypothesis, we employed a mixed approach combining quantitative user profiling with qualitative semi-structured interviews and task-based contextual experiments. Based on these interactions, we attempt to correlate reported behavior (what they say) with the actual interaction (what they do).

\subsection{Social Profile of Participants}

We recruited 16 participants who are active users of platforms such as Zepto, Blinkit, and Instamart All participants were recruited through criterion-based purposive sampling, identifying young urban Indians who met the study's digital fluency and financial constraint criteria within the authors' extended social and professional networks.The selection reflected a digital-native demographic with high smartphone proficiency but varying levels of financial autonomy (Table \ref{tab:social-profile-updated}). Prior to the interview, participants completed a structured questionnaire to establish their financial baseline and usage habits. This phase was critical to segment users based on overall financial awareness rather than just income. The cities of the participants have been classified into tiers (1,2 or 3) based on the standard classification used by the Indian Government. The classification system categorises urban areas based on population density and other development metrics. We collected useful information about the user's financial constraints, their overall price sensitivity and usage across different platforms. This phase helps us define the baseline for each user, against which we can measure their actual behavior. While participants reported high price sensitivity (mean = 3.8/5), their behaviors during tasks often diverged from these stated preferences.

\subsection{Interviews}

We employed a semi-structured interview approach (approximately 30 minutes each), utilizing an interview guide that allowed for flexibility while ensuring core research questions were addressed. To move beyond mere descriptive narratives and better understand the `Awareness-Action Gap', we explicitly used `why' questions to probe the underlying rationales behind participants' behaviors and interactions with quick commerce interfaces. The questions that we planned on asking were from a few different themes such as:

\begin{itemize}

  \item \textbf{Usecase}: We probed usage scenarios to distinguish between genuine needs and manufactured urgencies (wants and cravings).

  \item \textbf{Trap Assessment}: Participants were asked to walk us through their reaction to common cues such as `Surge Pricing' or `High Demand'. We tried to analyze how they deal with these situations, keeping their user profile as a baseline for their claimed behavior.

  \item \textbf{Overall Awareness}: We explicitly assessed their knowledge of deceptive tactics and their perception of recent regulatory changes.

\end{itemize}

To observe the `Action' component of our hypothesis, we also conducted two small yet specific tasks with the participants on their preferred app, asking them to think out loud while doing so. This was done to allow us to capture real-time cognitive processing and their immediate reactions to the UI. We asked the participants to find out various things related to their subscription (such as purchase details, renewal, cancellation) and their account. We counted the number of clicks required and also the time taken by them to reach the relevant stages on the app. Also, we asked all participants to assemble a cart for a very small order (1 packet of milk and 1 packet of bread) and proceed to the final ordering screen. We did this to observe whether users are able to pay attention and acknowledge the breakdown of various added fees and charges.

The resulting data was analyzed using reflexive thematic analysis, following the systematic phases outlined by Braun and Clarke. This process involved an interpretive engagement with the data through both semantic coding (capturing explicit participant language) and latent coding (interpreting underlying assumptions and conceptual frameworks).

\begin{table}[h!]
  \centering
  \caption{Social Profile of Participants with City Tier Classification}
  \label{tab:social-profile-updated}
  \resizebox{\textwidth}{!}{%
    \begin{tabular}{|l|l|l|l|l|l|l|}
      \hline
      \textbf{ID} & \textbf{Age} & \textbf{City (Tier)} & \textbf{Primary Apps} & \textbf{Price Sensitivity (1-5)} & \textbf{Frequency} & \textbf{AOV (Rs.)} \\ \hline
      P1 & 22 & Kota (Tier 2) & Instamart, Blinkit & 4 & Weekly & 300--350 \\ \hline
      P2 & 21 & Delhi (Tier 1 - Metro) & Instamart, Zepto & 2 & 4--5x Weekly & 250--300 \\ \hline
      P3 & 22 & Delhi (Tier 1 - Metro) & Zepto, Instamart & 4 & 2--3x Weekly & 150--200 \\ \hline
      P4 & 20 & Surat (Tier 2) & Instamart, Zepto & 3 & Weekly & 150--200 \\ \hline
      P5 & 20 & Patiala (Tier 2) & Instamart, Zepto, Blinkit & 2 & 5x Weekly & $<$200 \\ \hline
      P6 & 20 & Delhi (Tier 1 - Metro) & Instamart, Blinkit & 5 & Bi-weekly & 200--300 \\ \hline
      P7 & 20 & Patna (Tier 2) & Instamart, Blinkit & 2 & Monthly & 200--300 \\ \hline
      P8 & 21 & Chandigarh (Tier 2) & Instamart, Zepto & 3 & Daily & 200--250 \\ \hline
      P9 & 18 & Delhi (Tier 1 - Metro) & Blinkit & 3 & Monthly & 500--600 \\ \hline
      P10 & 19 & Hyderabad (Tier 1 - Metro) & Instamart, Blinkit, Zepto & 5 & Bi-weekly & 200--300 \\ \hline
      P11 & 19 & Hyderabad (Tier 1 - Metro) & Zepto, Blinkit & 4 & Weekly & 200--300 \\ \hline
      P12 & 19 & Mumbai (Tier 1 - Metro) & Blinkit, Zepto & 4 & Bi-weekly & 300--350 \\ \hline
      P13 & 21 & Hyderabad (Tier 1 - Metro) & Blinkit, Zepto & 4 & Weekly & 300--350 \\ \hline
      P14 & 18 & Hyderabad (Tier 1 - Metro) & Zepto, Instamart & 5 & Monthly & 150--200 \\ \hline
      P15 & 19 & Meerut (Tier 3) & Zepto, Instamart & 4 & Every 10 Days & 150--200 \\ \hline
      P16 & 18 & Vadodara (Tier 2) & Zepto, Instamart & 4 & 2x Weekly & 100 \\ \hline
    \end{tabular}%
  }
\end{table}

\subsection{Research Ethics}
Adhering to ethical guidelines, we implemented a participant-centered method prioritizing comfort and agency. Each participant was briefed on the study's objectives and the interview protocol to secure informed and explicit consent. To protect privacy, all participants were anonymized as P1 through P16, and specific identifying details were omitted. All data was securely stored and accessed solely by the authors to prevent unauthorized exposure.

\section{Findings}

Our qualitative analysis reveals a pervasive Awareness-Action Gap among Indian university students, where the structural design of Quick Commerce platforms reinforces manipulative practices that are accepted as standard industry behavior. We identify thematic pillars that define this interaction: the opportunistic acquisition of memberships through nominal friction, the weaponization of temporal urgency via gamification, and the normalization of filler consumption to bypass interface interference.

The first major finding identifies an enrollment pattern where memberships are acquired through low-cost nudges at the checkout stage rather than deliberate long-term planning. Several participants reported purchasing memberships for nominal fees, often ranging from 1 to 30 rupees, specifically to offset delivery charges on a single immediate order. P10 described this as a spontaneous reaction to the checkout interface, stating that ``whenever I add items to my cart, there is an option to include a membership for just one rupee, which reduces the total price of the bulk order''. This nominal entry creates a psychological bypass where the low initial cost masks the complexity of renewal. In some cases, enrollment was perceived as entirely unintentional. P13 noted that the membership was added to his cart automatically without explicit prior intent, stating that ``the membership was entirely unplanned; the app added it to my cart automatically and the transaction proceeded from there''.

This initial ease of entry is contrasted by significant downstream friction, embodying the Roach Motel pattern. When participants were tasked with navigating cancellation paths during our contextual audit, the systemic obscuring of exit options became starkly evident. P3 remarked that she ``had no idea how to cancel the subscription because even after looking through the account and membership settings, the option to deactivate is simply not visible''. The frustration regarding these deceptive architectures was high enough that participants suggested extreme measures to regain autonomy. P12 suggested that the only way to escape the cycle might be through measures such as ``uninstalling the app entirely or replacing contact information with a dummy phone number'' because a clear deactivation option was not provided.

The second theme centers on the use of manufactured urgency and gamification to override the price sensitivity of participants. Temporal dark patterns were found to be significantly amplified during high-engagement cultural events. P11 recalled how specific events in sports triggered discount windows that induced higher spending, explaining that ``during events like the World Cup, the app prompts you to place an order whenever a six is hit to receive a 66 percent discount, which definitely encourages more frequent ordering''. This sense of urgency creates a cognitive load that prevents the price-comparison behavior usually practiced by this demographic. P10 confirmed this behavioral shift, noting that during the IPL, short timers for ordering food caused him to ``change my pace and act more quickly''.

Even outside of specific gamified events, High Demand messaging acts as a barrier that increases platform lock-in. P14 described the ambiguity of being blacked out late at night, where the app would prevent ordering for extended periods, stating that ``the app restricts ordering for thirty to forty minutes during peak times, and it remains ambiguous whether or not a transaction will eventually be permitted''. Rather than switching to a physical alternative, users often chose to wait for the block to lift, demonstrating that the perception of urgency creates a state where agency is traded for eventual convenience. P13 noted that when seeing a timer on a deal, he ``finds himself repeatedly checking the countdown, debating whether to secure the current deal or continue searching for the item originally intended''.

The Small Cart dilemma further illustrates the gap between awareness and action. In our task-based contextual experiment, a 32 rupee packet of milk frequently escalated to nearly 100 rupees or more at checkout due to hidden fees. P14 observed the absurdity of this pricing structure during the live test, noting that ``a simple order starting at 32 rupees eventually escalated to a total of 132 rupees, yet the interface then claims a saving of five rupees, which feels entirely deceptive''. To mitigate these costs, participants frequently engage in filler consumption. P9 explained that because of free delivery limits, he feels the need to add more items to the cart, such as ``chocolates or chips, simply to reach the free delivery threshold''. This creates a sense of false gratification that bypasses the critical scrutiny of users who otherwise claim to be highly price-sensitive.

Confusing defaults regarding ancillary items provided another layer of interface interference. P7 recounted an experience where a cutlery checkbox was structured so non-intuitively that it resulted in a complete lack of utility, describing the wording as ``the most frustrating design I have encountered, which often confuses even the restaurants''. He further emphasized the real-world consequence, noting he ``received the order without utensils and had to eat with his hands''. The seamless nature of the payment flow further inhibits rational decision-making. P10 remarked that ``the transaction process is so seamless that one can simply slide to pay immediately, and within that rapid flow, these underlying patterns often go unnoticed''.

Ultimately, our findings suggest that the Awareness-Action Gap is a calculated byproduct of Digital Resignation. Participants frequently used terms like ``seamless'' to justify their lack of scrutiny, suggesting that the speed of Q-Commerce acts as a cognitive buffer. P14 characterized deceptive design as an inescapable systemic issue, stating that ``these deceptive designs are essentially the price one pays for living under modern capitalism''. This demographic has internalized dark patterns as a foundational element of the digital economy. The fact that users still struggle to find clear exit paths indicates that platforms are engaging in malicious compliance. P12 explicitly stated his belief regarding the lack of oversight, noting that he ``suspects these platforms are not being monitored by the government at all, allowing these practices to continue unchecked''.

\section{Discussion} The findings suggest that the `Awareness-Action Gap' is a calculated byproduct of `Seamless Design' and `Digital Resignation'. For the users, convenience has been commodified to the point where it overrides financial rationality. Participants frequently used terms like ``seamless'' to justify their lack of scrutiny, suggesting that the extreme speed of Quick Commerce acts as a cognitive buffer that prevents users from engaging in analytical thinking. This creates an environment where the perceived utility of 10-minute delivery makes the erosion of autonomy an acceptable trade-off.

We also identify a state of `Digital Resignation', where users recognize they are being manipulated but feel that resistance is futile. P14's characterization of deceptive design as the ``price of capitalism'' and P11's view that it is ``just business'' reflect a demographic that has internalized dark patterns as a foundational element of the digital economy. This resignation is a significant barrier to ethical design; if users do not expect or demand transparency, there is little market incentive for service providers to offer it, leading to a race to the bottom in terms of UX ethics.

The gap between the CCPA Guidelines and what participants actually experienced shows up in specific, named provisions. The Guidelines explicitly prohibit \textbf{Drip Pricing}, defined as revealing charges incrementally rather than upfront. Every participant who completed the milk-and-bread task encountered exactly this: a Rs 32 item ballooning to Rs 97 or Rs 132 at checkout, with handling fees, small-cart charges, and surge fees appearing only at the final screen. P14's observation that the app then claimed she was ``saving five rupees'' on a Rs 132 order she hadn't intended to place captures how this provision is being violated in spirit while remaining arguable on paper.

\textbf{Basket Sneaking}, which involves the automatic addition of items without explicit consent, appeared in at least two accounts: P13 described Zepto adding a membership ``to my cart for me,'' and a participant in the interview cohort described the same for Swiggy One.

The Guidelines also specifically prohibit \textbf{Subscription Traps}, requiring that cancellation be as accessible as signup. Yet no participant who attempted the cancellation task found a direct path to it; several concluded the only exit was uninstalling the app or switching to a dummy phone number.

The cutlery checkbox that P7 described is a double-negative opt-out that caused him to eat noodles without cutlery, which falls squarely under \textbf{Trick Questions}, another listed provision. That not a single participant among the sixteen was aware that any of this had been banned points to a compliance theatre problem rather than a genuine regulatory one.

To align commercial incentives with ethical practice, we propose a shift from `Convenience at all costs' to `Informed Flow'. This would require mandating equal exit paths, where a cancellation button is as visually prominent as a Buy button, and ensuring real-time fee visualization on product pages rather than surprising users at the final stage of a transaction. By re-introducing a `healthy friction' into the user journey, designers can help bridge the gap between user expectations and the commercial pressures of the Quick Commerce model.

\section{Conclusion}
This study confirms that Indian university students are acutely aware of deceptive design but are structurally nudged into succumbing to it due to induced urgency and a perceived lack of alternatives. The awareness-action gap is facilitated by the perceived immense utility of near-instant delivery, which transforms convenience from a luxury into a necessity. Regulatory efforts must move beyond defining patterns to mandating fair exit paths and proactive disclosures. Future research should investigate if these behaviors differ across age groups or if the high-pressure environment of student life makes this demographic uniquely vulnerable to the tradeoff between transparency and convenience.

\section*{Declaration on Generative AI}
 During the preparation of this work, the authors used Gemini 3 for: Grammar and Spelling check along with Citation Management. After using the service, the authors reviewed and edited the content as needed and take full responsibility for the publication's content.

\bibliography{sample-ceur}

@misc{brignull-website,
author = {Brignull, H and Leiser, M and Santos, C and Doshi, K},
month = {4},
title = {{Deceptive patterns – user interfaces designed to trick you}},
year = {2023},
url = {https://www.deceptive.design/},
}

@mastersthesis{juneja2025,
  author  = {Riya Juneja},
  title   = {Dark Patterns in Indian Digital Interfaces: Prevalence, Cultural Perception and User Education},
  school  = {Purdue University},
  year    = {2025},
  note    = {Master of Science Thesis}
}

@article{dark-patterns-at-scale,
   title={Dark Patterns at Scale: Findings from a Crawl of 11K Shopping Websites},
   volume={3},
   ISSN={2573-0142},
   url={http://dx.doi.org/10.1145/3359183},
   DOI={10.1145/3359183},
   number={CSCW},
   journal={Proceedings of the ACM on Human-Computer Interaction},
   year={2019},
   publisher={Association for Computing Machinery (ACM)},
   author={Mathur, Arunesh and Acar, Gunes and Friedman, Michael J. and Lucherini, Eli and Mayer, Jonathan and Chetty, Marshini and Narayanan, Arvind},
}

@article{chugh2024unpacking,
  title={Unpacking Dark Patterns: Understanding Dark Patterns and Their Implications for Consumer Protection in the Digital Economy},
  author={Chugh, B and Jain, P},
  journal={Consumer Protection Review},
  year={2024}
}

@article{raj2025safeguarding,
  title={Safeguarding the Digital Consumer: A Comparative Legal and Psychological Analysis of Dark Patterns in E-Commerce},
  author={Raj, P and Nanda, S S and Noorani, M S},
  journal={Advances in Consumer Research},
  year={2025}
}

@article{awasthi2025dark,
  title={The Dark Pattern Free Future of E-Commerce in India},
  author={Awasthi, S and Soni, Y},
  journal={Shodh Samagam},
  year={2025}
}

@article{kumar2025dark,
  title={Dark Pattern in Online Trading - A Critical Examination of their Legality under Consumer Protection Laws},
  author={Kumar, D P and Krishna, K S},
  journal={Gurukul International Multidisciplinary Research Journal},
  year={2025}
}

@article{qcom-gupta,
  title = {A Study on Emergence of Quick Commerce},
  volume = {6},
  ISSN = {2582-2160},
  url = {http://dx.doi.org/10.36948/ijfmr.2024.v06i02.19226},
  DOI = {10.36948/ijfmr.2024.v06i02.19226},
  number = {2},
  journal = {International Journal For Multidisciplinary Research},
  publisher = {International Journal for Multidisciplinary Research (IJFMR)},
  author = {Shivom Gupta},
  year = {2024},
  month = apr 
}

@article{qcom-singh,
  title = {The Impact of Quick Commerce on Consumer Behavior and Economic Trends in India: A Systematic Review},
  volume = {09},
  ISSN = {2455-8834},
  url = {http://dx.doi.org/10.46609/IJSSER.2024.v09i08.020},
  DOI = {10.46609/ijsser.2024.v09i08.020},
  number = {08},
  journal = {International Journal of Social Science and Economic Research},
  publisher = {MIJP Publication},
  author = {Singh,  Risbaa},
  year = {2024},
  pages = {2859–2874}
}

@article{qcom-goyal,
  title = {Key Factors Driving the Rapid Growth of Quick Commerce in Urban Areas of India},
  volume = {6},
  ISSN = {2582-2160},
  url = {http://dx.doi.org/10.36948/ijfmr.2024.v06i06.31781},
  DOI = {10.36948/ijfmr.2024.v06i06.31781},
  number = {6},
  journal = {International Journal For Multidisciplinary Research},
  publisher = {International Journal for Multidisciplinary Research (IJFMR)},
  author = {Anshika Goyal},
  year = {2024},
  month = nov 
}

@article{qcom-ganpathy,
  title = {Critical success factors for quick commerce grocery delivery in India: an exploratory study.},
  volume = {12},
  ISSN = {0719-3726},
  url = {http://dx.doi.org/10.7770/safer-V12N1-art691},
  DOI = {10.7770/safer-v12n1-art691},
  number = {1},
  journal = {Sustainability,  Agri,  Food and Environmental Research-DISCONTINUED},
  publisher = {Universidad Catolica de Temuco},
  author = {Ganapathy,  Venkatesh and Dr Chithambar Gupta},
  year = {2023},
  month = sep 
}

@misc{ccpa2023,
  author = {{Central Consumer Protection Authority}},
  title  = {Guidelines for Prevention and Regulation of Dark Patterns},
  year   = {2023},
  publisher = {Government of India},
  note   = {Classifies 13 deceptive practices as unfair trade}
}

@inproceedings{gray2018,
  author = {Colin M. Gray and Yubo Kou and Bryan Battles and Joseph Hoggatt and Austin L. Toombs},
  title = {The Dark (Patterns) Side of UX Design},
  booktitle = {Proceedings of the 2018 CHI Conference on Human Factors in Computing Systems},
  year = {2018},
  publisher = {ACM}
}

@inproceedings{gray2024, series={CHI ’24},
   title={An Ontology of Dark Patterns Knowledge: Foundations, Definitions, and a Pathway for Shared Knowledge-Building},
   url={http://dx.doi.org/10.1145/3613904.3642436},
   DOI={10.1145/3613904.3642436},
   booktitle={Proceedings of the CHI Conference on Human Factors in Computing Systems},
   publisher={ACM},
   author={Gray, Colin M. and Santos, Cristiana Teixeira and Bielova, Nataliia and Mildner, Thomas},
   year={2024},
   month=may, pages={1–22},
   collection={CHI ’24} }

@article{santos-legal,
  title={No harm no foul: how harms caused by dark patterns are conceptualised and tackled under EU data protection, consumer and competition laws},
  author={Cristiana Santos and Viktorija Morozovaite and Silvia De Conca},
  journal={Information \& Communications Technology Law},
  year={2025},
  volume={34},
  pages={329 - 375},
  url={https://api.semanticscholar.org/CorpusID:276457124}
}

@article{mamidwar-legal,
  author = {Aryan Mamidwar and Ganesh Bhutkar},
  title = {An Overview of Guidelines on Dark Patterns},
  journal = {CHI’24: Mobilizing Research and Regulatory Action on Dark Patterns and Deceptive Design Practices},
  year = {2024}
}

@article{chen2023unveiling,
  title={Unveiling the Tricks: Automated Detection of Dark Patterns in Mobile Applications},
  author={Chen, J and Sun, J and Feng, S and others},
  journal={arXiv preprint arXiv:2308.05898},
  year={2023}
}

@inproceedings{gray2025iliad,
  title={Getting Trapped in Amazon's ``Iliad Flow'': A Foundation for the Temporal Analysis of Dark Patterns},
  author={Gray, Colin M and Mildner, Thomas and Gairola, Ritika},
  booktitle={Proceedings of the CHI Conference on Human Factors in Computing Systems (CHI '25)},
  year={2025},
  publisher={ACM},
  address={Yokohama, Japan},
  doi={10.1145/3706598.3713828}
}

@article{wu2022malicious,
  title={Malicious Selling Strategies in E-Commerce Livestream: A Case Study},
  author={Wu, Qunfang and Sang, Yisi and Wang, D and Lu, Zhicong},
  journal={arXiv preprint arXiv:2201.00000},
  note={See also: Proceedings of CHI 2022 or CSCW if published},
  year={2022}
}

@misc{chen-2024,
  doi = {10.48550/ARXIV.2411.18084},
  url = {https://arxiv.org/abs/2411.18084},
  author = {Chen,  Jieshan and Wang,  Zhen and Sun,  Jiamou and Zou,  Wenbo and Xing,  Zhenchang and Lu,  Qinghua and Huang,  Qing and Xu,  Xiwei},
  title = {From Exploration to Revelation: Detecting Dark Patterns in Mobile Apps},
  publisher = {arXiv},
  year = {2024},
  copyright = {Creative Commons Attribution Non Commercial No Derivatives 4.0 International}
}

@book{thaler2008nudge,
  title={Nudge: Improving decisions about health, wealth, and happiness},
  author={Thaler, Richard H and Sunstein, Cass R},
  year={2008},
  publisher={Yale University Press}
}

@article{luguri2021shining,
  title={Shining a light on dark patterns},
  author={Luguri, Jamie and Strahilevitz, Lior Jacob},
  journal={Journal of Legal Analysis},
  volume={13},
  number={1},
  pages={43--109},
  year={2021},
  publisher={Oxford University Press},
  doi={10.1093/jla/laaa006}
}

@article{Narayanan2020DarkPP,
  title={Dark Patterns: Past, Present, and Future},
  author={Arvind Narayanan and Arunesh Mathur and Marshini Chetty and Mihir Kshirsagar},
  journal={Queue},
  year={2020},
  volume={18},
  pages={67 - 92},
  url={https://api.semanticscholar.org/CorpusID:219666267}
}

@inproceedings{bongard2021, series={DIS ’21},
   title={”I am Definitely Manipulated, Even When I am Aware of it. It’s Ridiculous!” - Dark Patterns from the End-User Perspective},
   url={http://dx.doi.org/10.1145/3461778.3462086},
   DOI={10.1145/3461778.3462086},
   booktitle={Designing Interactive Systems Conference 2021},
   publisher={ACM},
   author={Bongard-Blanchy, Kerstin and Rossi, Arianna and Rivas, Salvador and Doublet, Sophie and Koenig, Vincent and Lenzini, Gabriele},
   year={2021},
   month=jun, pages={763–776},
   collection={DIS ’21} }

@article{chen-etal,
  title={From Exploration to Revelation: Detecting Dark Patterns in Mobile Apps},
  author={Jieshan Chen and Zhen Wang and Jiamou Sun and Wenbo Zou and Zhenchang Xing and Qinghua Lu and Qing Huang and Xiwei Xu},
  journal={ArXiv},
  year={2024},
  volume={abs/2411.18084},
  url={https://api.semanticscholar.org/CorpusID:274306395}
}

@INPROCEEDINGS{geronimo-etal,
  title      = "{UI} dark patterns and where to find them",
  booktitle  = "Proceedings of the 2020 {CHI} Conference on Human Factors in
                Computing Systems",
  author     = "Di Geronimo, Linda and Braz, Larissa and Fregnan, Enrico and
                Palomba, Fabio and Bacchelli, Alberto",
  publisher  = "ACM",
  pages      = "1--14",
  month      =  apr,
  year       =  2020,
  address    = "New York, NY, USA",
  conference = "CHI '20: CHI Conference on Human Factors in Computing Systems",
  location   = "Honolulu HI USA"
}

@ARTICLE{gunawan-etal,
  title     = "A comparative study of dark patterns across web and mobile
               modalities",
  author    = "Gunawan, Johanna and Pradeep, Amogh and Choffnes, David and
               Hartzog, Woodrow and Wilson, Christo",
  journal   = "Proc. ACM Hum. Comput. Interact.",
  publisher = "Association for Computing Machinery (ACM)",
  volume    =  5,
  number    = "CSCW2",
  pages     = "1--29",
  month     =  oct,
  year      =  2021,
  copyright = "http://www.acm.org/publications/policies/copyright\_policy\#Background",
  language  = "en"
}
\end{document}